\documentclass[prl,twocolumn,superscriptaddress,preprintnumbers]{revtex4}
\usepackage{amsfonts}
\usepackage{amssymb}
\usepackage{amsmath}
\usepackage{color}
\definecolor{darkblue}{rgb}{0.15,0.35,0.55}
\definecolor{reddish}{rgb}{0.65, 0.2, 0.2}
\usepackage[bookmarks,linktocpage, colorlinks=true, plainpages = false, citecolor = darkblue,  linkcolor=darkblue, urlcolor = darkblue, filecolor = darkblue]{hyperref} 

\begin{document}
\preprint{YITP-21-128, IPMU21-0072}

\title{DeWitt boundary condition is consistent in Ho\v{r}ava-Lifshitz quantum gravity}

\author{Hiroki Matsui}
\email{hiroki.matsui@yukawa.kyoto-u.ac.jp}
\affiliation{Center for Gravitational Physics and Quantum Information, Yukawa Institute for Theoretical Physics, Kyoto University, 606-8502, Kyoto, Japan}
\author{Shinji Mukohyama}
\email{shinji.mukohyama@yukawa.kyoto-u.ac.jp}
\affiliation{Center for Gravitational Physics, Yukawa Institute for Theoretical Physics, Kyoto University, 606-8502, Kyoto, Japan}
\affiliation{Kavli Institute for the Physics and Mathematics of the Universe (WPI), The University of Tokyo Institutes for Advanced Study, The University of Tokyo, Kashiwa, Chiba 277-8583, Japan}
\author{Atsushi Naruko}
\email{naruko@yukawa.kyoto-u.ac.jp}
\affiliation{Center for Gravitational Physics, Yukawa Institute for Theoretical Physics, Kyoto University, 606-8502, Kyoto, Japan}

\begin{abstract}
In quantum cosmology the DeWitt boundary condition is a proposal to set the wave function of the universe to vanish at the classical big-bang singularity. In this Letter, we show that in many gravitational theories including general relativity, the DeWitt wave function does not take a desired form once tensor perturbations around a homogeneous and isotropic closed universe are taken into account: anisotropies and inhomogeneities due to the perturbations are not suppressed near the classical singularity. We then show that Ho\v{r}ava-Lifshitz gravity provides a satisfactory DeWitt wave function. In particular, in the limit of $z=3$ anisotropic scaling, we find an exact analytic expression for the DeWitt wave function of the universe with scale-invariant perturbations. In general cases with relevant deformations, we show that the DeWitt wave function can be systematically expanded around the classical big-bang singularity with perturbations under control. 
\end{abstract}
\date{\today}
\maketitle

\section{Introduction}
\label{sec:intro} 

Quantum cosmology is an attempt to describe the entire universe based on quantum theory. Due to the lack of a complete theory of quantum gravity, however, the quantization of spacetime is not a trivial task, especially in general relativity (GR). Nonetheless there exist several approaches to quantum cosmology, in which one hopes to grasp a coarse-grained description of the quantized spacetime in the very early universe. 

One of them is based on the canonical quantum gravity~\cite{DeWitt:1967yk}, which treats gravity in the Hamiltonian formulation using the Arnowitt, Deser and Misner (ADM) formalism~\cite{Arnowitt:1962hi}. In this approach the Hamiltonian constraint is interpreted as an operator equation, $\hat{H}[g]\Psi[g]=0$, called the Wheeler-DeWitt equation. Here, $g$ represents the spatial metric induced on a $3$-geometry, $\hat{H}[g]$ is the operator corresponding to the Hamiltonian constraint and $\Psi[g]$ is the so-called wave function of the universe. Alternatively, the wave function of the universe can be formulated by the path integral, $\Psi[g]=\int\mathcal{D} g^{(4)}\, e^{i S[g^{(4)}] / \hbar}$, where it is understood that the $4$-dimensional metric $g^{(4)}$ is restricted to those inducing $g$ on the $3$-geometry and that the diffeomorphism invariance is properly treated~\cite{Halliwell:1988wc}. The Wheeler-DeWitt equation is thought to describe a coarse-grained nature of the entire quantum universe but allows for various solutions. Therefore, in order to select the wave function of the universe, a proper boundary condition has to be imposed. Correspondingly, in the path integral approach one needs to specify the range and contour of the path integral.

In quantum cosmology, there are two famous boundary conditions to define the wave function of the universe: the no-boundary proposal~\cite{Hartle:1983ai} and the tunneling proposal~\cite{Vilenkin:1984wp}. However, there are some doubts on these proposals under the inclusion of perturbations around a homogeneous and isotropic background~\cite{Feldbrugge:2017fcc,Feldbrugge:2017mbc}. Based on the real-time path integral formulation, it has been claimed that the wave function of small perturbations around the background takes the form of inverse-Gaussian and will be out of control. For instance, when tensor perturbation $h$ is included, the wave function takes the form $\Psi(a,h)\propto \exp[{+\alpha(k,a)h_k^2/ \hbar}]$ for a mode $h_k$ with the comoving wavenumber $k$, where $\alpha(k,a)>0$. A similar conclusion holds for scalar perturbations as well.  These results are likely to be inconsistent with cosmological observations. The issue of perturbations in the no-boundary and tunneling proposals has recently been further discussed in the literature~\cite{DiazDorronsoro:2018wro,Feldbrugge:2018gin,Vilenkin:2018dch, Bojowald:2018gdt,Halliwell:2018ejl,Vilenkin:2018oja,DiTucci:2019dji,Wang:2019spw,
DiTucci:2019xcr,Lehners:2021jmv}.

In this Letter, we instead adopt the so-called {\it DeWitt boundary condition}, which states that the wave function of the universe should vanish at the classical big-bang singularity~\cite{DeWitt:1967yk}. For a homogeneous and isotropic universe the DeWitt boundary condition can be expressed as $\Psi(a=0)=0$, which is known to successfully regularize the behavior of the wave function near the classical singularity. In the minisuperspace  where the dynamics of the universe are only parameterized by the scale factor $a(t)$,  one can easily find an analytic expression for the {\it DeWitt wave function}, i.e. the solution to the Wheeler-DeWitt equation with the DeWitt boundary condition. However, the generalization beyond the minisuperspace is not trivial. In this Letter, we actually show that in many gravity theories including GR, the DeWitt wave function for a homogeneous and isotropic background with small  perturbations is not well-behaved and that the supposedly small perturbations cannot be suppressed near the classical big-bang singularity.

This is a serious problem in gravity theories including GR if one is to adopt the DeWitt wave function as a description of the very early universe. This suggests that the DeWitt boundary condition introduced as a proposal to tame the classical big-bang singularity in quantum cosmology requires gravity beyond GR. Fortunately, in the context of Ho\v{r}ava-Lifshitz (HL) gravity~\cite{Horava:2009uw}, we find that the introduction of higher dimensional operators that are required by perturbative renormalizability renders the wave function of perturbations well-behaved. Indeed, the wave function is shown to be of the form of a Gaussian distribution for the vacuum of the perturbations (or similarly suppressed distributions for excited states) all the way up to the classical big-bang singularity. This is a reminiscence of the fact that the same higher dimensional operators lead to a novel generation mechanism of scale-invariant cosmological perturbations without inflation~\cite{Mukohyama:2009gg}.

The rest of the Letter is organized as follows. First we show a No-go result in GR for the DeWitt wave function: tensor perturbations around a homogeneous and isotropic closed universe are not suppressed at the classical big-bang singularity. Then we extend the theory of gravity and show that the Ho\v{r}ava-Lifshitz gravity provides a satisfactory DeWitt wave function. Thus the DeWitt boundary condition is consistent in the Ho\v{r}ava-Lifshitz quantum gravity.

\section{No-go in GR}

We shall begin with the analysis in GR and show a negative result.
The gravitational action $S_{\rm GR}$ is written as
\begin{equation}
S_{\rm GR} = \frac{M_{\rm Pl}^2}{2}\int dt d^3 x N \sqrt{g} \left(K^{ij}K_{ij}- K^2 +R-2\Lambda\right)\,. 
\end{equation}
with the ADM form of the metric~\cite{Arnowitt:1962hi}, 
$d s^2 = - N^2 d t^2 + g_{i j} (d x^i + N^i dt) (d x^j + N^j dt)$,
where $N$, $N ^i$ and $g_{i j}$ are respectively the lapse function, the shift vector and the $3$-dimensional spatial metric. $K_{ij}$ is the extrinsic curvature tensor defined by
$K_{ij}= (\partial_t g_{ij}- g_{j k}\nabla_iN^k-g_{i k}\nabla_jN^k)/(2N)$,
 with $\nabla_i$ being the spatial covariant derivative compatible with $g_{ij}$, $K^{ij}=g^{ik}g^{jl}K_{kl}$ and $K=g^{ij}K_{ij}$ where $g^{ij}$ is the inverse of $g_{ij}$. $M_{\rm Pl}=1/\sqrt{8\pi G}$ is the Planck mass, $R$ is the Ricci scalar of $g_{ij}$, and $\Lambda$ is the cosmological constant. Hereafter we adopt the unit with $M_{\rm Pl}=1$.

To simplify the analysis we consider a closed Friedmann-Lema\^{i}tre-Robertson-Walker (FLRW) universe with tensor-type metric perturbations. The metric takes 
$d s^{2}=-N^{2}(t) d t^{2}+a^{2}(t)\left[\Omega_{ij} ({\bf x})+h_{i j} (t \,, {\bf x}) \right] d x^{i} d x^{j}$,
where $\Omega_{ij}$ is the metric of the unit 3-sphere~\footnote{The corresponding Riemann curvature is $R^{ij}_{kl}[\Omega]=(\delta^i_k\delta^j_l-\delta^i_l\delta^j_k)$.} and $\Omega^{ij}$ is the inverse of $\Omega_{ij}$. $h_{ij}$ is the tensor perturbation satisfying the transverse and traceless condition, namely
 $\Omega ^{i j} h_{i j} = \Omega^{j k}D_k h_{i j} = 0$ where $D_i$ is the spatial covariant derivative compatible with $\Omega_{ij}$. Hereinafter, spatial indices $i, j, \cdots$ are raised and lowered by $\Omega^{ij}$ and $\Omega_{ij}$. For this metric, the action is expanded up to the second order in perturbation as $S_{\rm GR} = S_{\rm GR}^{(0)}+S_{\rm GR}^{(2)}+\mathcal{O}(h^3)$:
\begin{align}
\begin{split}\label{eq:GR}
S_{\rm GR}^{(0)} &=  V\int dt\, 
\left(\frac{3N}{a}\right)\biggl[
-\left(\frac{a}{N}\dot{a}\right)^{2}+
a^2-\frac{\Lambda a^4}{3} \biggr]\,, \\
S_{\rm GR}^{(2)} &=  \int dt \left({Na}\right) \int d^3x\sqrt{\Omega} \\
&\qquad\times \frac{1}{8} \biggl[
 \frac{a^2}{N^2}\dot{h}^{ij}\dot{h}_{ij}
 - h^{ij}\left(D^2-6 \right)h_{ij}\biggr]\,, 
\end{split}
\end{align}
 where $V=\int d^3x\sqrt{\Omega}=2\pi^2$ is the volume of the unit $3$-sphere and $D^2 = \Omega^{i j} D_i D_j$.

The tensor perturbation $h_{ij}$ can be expanded in terms of the tensor hyper-spherical harmonics~\cite{Gerlach:1978gy}, 
\begin{equation}\label{eq:perturbation}
h_{ij}(t,x^i)= \sum_{snlm} h^{s}_{nlm}(t)Q^{snlm}_{ij}\,,
\end{equation}
where $s=\pm$ is the polarization label, the integers ($n$, $l$, $m$) run over the ranges $n\geq3$, $l \in [0,n-1]$, $m \in [-l,l]$, and $Q^{snlm}_{ij}$ are the tensor eigenfunctions of the Laplacian operator $D^2 [\Omega]$ on the unit $3$-sphere,
$D^2 [\Omega]\, Q^{snlm}_{ij}= -\left(n^2-3\right)Q^{snlm}_{ij}$,
normalized as 
$\int d^3x\sqrt{\Omega}\Omega^{ik}\Omega^{jl}Q^{snlm}_{ij}Q^{s'n'l'm'}_{kl} 
  = V\delta^{ss'}\delta^{nn'}\delta^{ll'}\delta^{mm'}$.

Substituting the expansion~(\ref{eq:perturbation}) into the action~(\ref{eq:GR}), we obtain
\begin{align}
&S_{\rm GR}^{(0)}+S_{\rm GR}^{(2)} =  V\int dt
\Biggl\{\left(\frac{3N}{a}\right)\biggl[
-\left(\frac{a}{N}\dot{a}\right)^{2}+
a^2-\frac{\Lambda a^4}{3} \biggr] \notag \\
&+ \sum_{snlm} \frac{1}{8} \left({Na}\right)\biggl[
 \left(\frac{a}{N}\dot{h}^s_{nlm}\right)^2
+ \left(n^2+3\right)(h^s_{nlm})^2\biggr]\Biggr\}\,.
\end{align}
Hereafter, for simplicity we restrict our consideration to the dynamics of the scale factor and one mode of the tensor perturbation. We then denote $h^s_{nlm}$ of our interest by $h$, suppressing the indices $snlm$.

Following the standard canonical quantization procedure, where the canonical momenta $\Pi_{a}$ and $\Pi_{h}$ conjugate respectively to $a$ and $h$ are transformed to Hermitian operators $-i\partial /\partial a$ and $-i\partial /\partial h$, the Hamiltonian constraint of the system is transformed to the Wheeler-DeWitt equation, 
\begin{align}
 &\left\{
  \frac{1}{2\gamma}\left(\frac{\partial^2}{\partial a^2} + \frac{p}{a}\frac{\partial}{\partial a}\right)
  + \left( - 3 a^2 + \Lambda a^4\right)\right. \nonumber\\
 &\quad\left. 
 - \left[\frac{2}{a^2V^2}\frac{\partial^2}{\partial h^2}
 +\frac{a^2}{8}\left(n^2+3 \right)h^2\right]\right\}
 \Psi(a,h) = 0\,, \label{eqn:WDWeq-GR}
\end{align}
where $\Psi(a,h)$ is the wave function of the universe.
Here, we have defined $\gamma=6V^2$ and introduced the parameter $p$ in order to take into account the ambiguity of the operator ordering. In quantum cosmology there are two well-known choices of $p$: the Laplace-Beltrami operator ordering ($p=1$) and the Vilenkin operator ordering ($p=-1$)~\cite{Hawking:1985bk,Vilenkin:1987kf,Steigl:2005fk}. To maintain the generality, however, we keep $p$ as an arbitrary constant.

To seek a solution of the equation (\ref{eqn:WDWeq-GR}) 
we employ the DeWitt boundary condition
$\Psi(0,h) = 0$ for ${}^{\forall}h$.
More specifically, we demand that
\begin{equation}
 \Psi(a,h) = a^c\sum_{i=0}^{\infty} F_i(h)\, a^i\,, \label{eqn:Psi-expanded}
\end{equation}
for small $a$, where $c$ is a positive constant and we assume that $F_0(h)$ is not identically zero. In addition, as a necessary condition for $\Psi(a,h)$ to give non-divergent correlation functions of $h$ on $a=const.$ hypersurfaces (with a reasonable choice of the norm that we do not need to specify)
~\footnote{
To construct the probability measure or inner product is an unsolved problem in quantum cosmology. However, there are several ways to avoid these problems, e.g. by considering the conserved current with taking an appropriate equal-time surfaces~\cite{DeWitt:1967yk} or the probability interpretation based on the Schr\"{o}dinger inner product~\cite{Wiltshire:1995vk}. },
we demand that 
\begin{equation}
 \lim_{h\to\pm\infty} F_i(h) = 0\,, \quad (i=0,1,\cdots)\,. 
  \label{eqn:bc-phi}
\end{equation}
Otherwise, correlation functions of $h$ (such as the power spectrum) on $a=const.$ hypersurfaces would diverge. In order to determine the positive constant $c$, we demand that $F_0(h)$ be a non-trivial smooth function satisfying the condition (\ref{eqn:bc-phi}) (with $i=0$) so that the leading behavior of $\Psi(a,h)$ near $a=0$ is $\Psi(a,h)\simeq a^cF_0(h) + O(a^{c+1})$.

By substituting (\ref{eqn:Psi-expanded}) to the Wheeler-DeWitt equation (\ref{eqn:WDWeq-GR}), at the leading order in $a$ we obtain
\begin{equation}
 \partial_{h}^2F_0 - V^2 (c+p-1)c \, F_0 = 0\,. \label{eqn:GR-leading-a}
\end{equation}
For any values of the parameters, there is no non-trivial smooth function $F_0(h)$ satisfying the condition (\ref{eqn:bc-phi}) (with $i=0$). (This conclusion holds even for a complex $c$.) In other words, no DeWitt wave function gives non-divergent correlation function of $h$ on $a=const.$ hypersurfaces near the classical big-bang singularity. Although we have shown this only for GR, this no-go result is quite generic and is applied  to other theories of gravity as long as higher spatial derivative terms are absent as we shall see below.

\section{Ho\v{r}ava-Lifshitz gravity}

In the following we shall show that the above no-go result can be avoided in the Ho\v{r}ava-Lifshitz (HL) gravity~\cite{Horava:2009uw}, which is renormalizable, unitary and regarded as one ultraviolet (UV) completion possibility of quantum gravity.

In the HL gravity the anisotropic scaling $(t,\vec{x}) \to (b^z t, b \vec{x})$ with the dynamical critical exponent $z=3$ in the ultraviolet (UV) regime ensures the renormalizability~\cite{Barvinsky:2015kil,Barvinsky:2017zlx}. In cosmology, this scaling has some intriguing implications. It serves as a mechanism of generating scale-invariant cosmological perturbations~\cite{Mukohyama:2009gg}, solving the horizon problem without inflation. It also provides the so-called  {\it anisotropic instanton}, which is expected to solve the flatness problem~\cite{Bramberger:2017tid}.

In the following, we consider the projectable HL gravity, where the lapse function is dependent only on time, $N=N(t)$. In the notation of \cite{Mukohyama:2010xz}, the action is given by
\begin{align}
\begin{split}
 S_{\rm HL}&= \frac{{\cal M}_{\rm HL}^2}{2}\int dt d^3\vec{x} N \sqrt{g} \,\Bigl(K^{ij}K_{ij}-\lambda K^2 +c^2_gR\\
&-2\Lambda+{\mathcal O}_{z>1} \Bigr)\,,
\end{split} 
\end{align}
where ${\cal M}_{\rm HL}$ is a mass scale and the higher dimensional operators ${\mathcal O}_{z>1}$ is given by
\begin{align}
\frac{{\mathcal O}_{z>1}}{2} & = 
 c_1\nabla_iR_{jk}\nabla^iR^{jk}+c_2\nabla_iR\nabla^iR+c_3R_i^jR_j^kR_k^i\nonumber\\
 &  +c_4RR_i^jR_j^i + c_5R^3 
  + c_6R_i^jR_j^i+c_7R^2 \,,
  \end{align}
$\lambda$ and $c_{n}$ ($n=1,\cdots,7$) are coupling constants that are subject to running under the renormalization group (RG) flow, $\nabla_{i}$ is the spatial covariant derivative compatible with the 3-metric $g_{ij}$. In the UV regime the terms with two time derivatives and those with six spatial derivatives are dominant, rendering $z=3$. On the other hand, in the infrared (IR) regime, higher derivative terms are not important and thus the theory automatically flows to $z=1$. If $\lambda$ flows to $1$ (from above) in the IR and if it does sufficiently quickly, then GR is recovered, thanks to an analogue of the Vainshtein mechanism~\cite{Mukohyama:2010xz,Izumi:2011eh,Gumrukcuoglu:2011ef}, 
and the linear instability of the scalar graviton does not show up~\cite{Mukohyama:2010xz}.

The $3$-dimensional space at each time may or may not be connected. To make the argument as general as possible we thus allow the $3$-dimensional space to be the union of connected pieces, $\Sigma_{\alpha}$ ($\alpha=1,\cdots$). In the following we call each $\Sigma_{\alpha}$ a local universe, while the union of all $\Sigma_{\alpha}$ represents the global universe. In this case, while the lapse function $N=N(t)$ is common for all $\alpha$, we have a set of shift vectors and a set of spatial metrics, $N^i = N_{\alpha}^i(t,\vec{x})$ and $g_{ij} = g^{\alpha}_{ij}(t,\vec{x})$ for $\vec{x}\in \Sigma_{\alpha}$.

Now we shall derive the Wheeler-DeWitt equation in the HL gravity. The wave function of the universe can be expressed as $\Psi = \prod_{\alpha}\Psi_{\alpha}(a_{\alpha}, h_{\alpha}; C_{\alpha})$ , where $a_\alpha$ stands for the scale factor of a local universe $\Sigma_{\alpha}$, $h_\alpha$ represents one mode of the tensor perturbations and $\{C_{\alpha}\}$ are separation constants satisfying $\sum_{\alpha}C_{\alpha}=0$. Each separation constant $C_{\alpha}$ corresponds to the amplitude of ``dark matter as integration constant''~\cite{Mukohyama:2009mz,Mukohyama:2009tp} in $\Sigma_{\alpha}$. A general solution can be then written as a linear combination of the special solutions, 
\begin{equation}
\Psi\left(
\{a_{\alpha}, h_{\alpha}\}\right)=\int \Biggl(\prod_{\beta}dC_{\beta}\Biggr)
A_{\{C_{\beta}\}}
\prod_{\alpha}\Psi_{\alpha}\left(a_{\alpha}, h_{\alpha}; C_{\alpha}\right)\,.
\end{equation}
In this expression, $\Psi\left(\{a_{\alpha}, h_{\alpha}\}\right)$ represents the wave function of the global universe while each $\Psi_{\alpha}\left(a_{\alpha}, h_{\alpha}; C_{\alpha}\right)$ represents the wave function of $\Sigma_{\alpha}$.

After the canonical quantization, we obtain  
\begin{align}
& \left\{
  \frac{1}{2}\left(\frac{\partial^2}{\partial a^2} + \frac{p}{a}\frac{\partial}{\partial a}\right)
  + \left( \mathcal{C} \, a - \frac{g_3}{a^2} - 3g_2 - 3g_1a^2 + g_0 a^4\right)\right. \nonumber\\
 & \left.
  -\frac{1}{2{V}^2a^2}\frac{\partial^2}{\partial\mathfrak{h}^2} 
 + \frac{\mathfrak{h}^2}{2}\left( f_1 a^2+ f_2 + \frac{f_3}{a^2}\right) \right\}
 \Psi(a,\mathfrak{h}) = 0\,. \label{eqn:WDWeq}
\end{align}
where we have omitted the index $\alpha$ and also abbreviated $\Psi_{\alpha}\left(a_{\alpha}, h_{\alpha}; C_{\alpha}\right)$ to $\Psi(a,\mathfrak{h})$. 
The above variable and parameters are defined by 
\begin{align}
&\mathfrak{h} = h_{\alpha}/(2\sqrt{\gamma}), \
\mathcal{C} = \gamma \, C_{\alpha}, \ g_3 = 24 \gamma (c_3+3c_4+9c_5),\notag \\
&g_2 = 4 \gamma (c_6+3c_7), \
g_1 = \gamma c_{\rm g}^2, \
g_0 = \gamma \Lambda, \notag \\
& f_1 = -\gamma^2 (n^2+3),
f_2 = - 8 \gamma^2 \Bigl[ c_6 (n^2+3)^2 + 18c_7(n^2-4) \Bigr] \notag \\
& f_3= - 8 \gamma^2 \Bigl[ - c_1 (n^6-9 n^4-9 n^2+81) + 6c_3 (n^2+3)^2 \notag
\\ &+6c_4 (n^4+9 n^2-3)+162 c_5 (n^2-4) \Bigr]
\end{align}
with $\gamma=3\left(3\lambda-1\right)V^2$.  This equation is applicable to both projectable and non-projectable HL theories, if one sets $\mathcal{C}=0$ for the latter.

\section{Scale-invariant solution}

Let us consider the special case where terms with $z < 3$ are absent. We set the parameters as 
$g_2 = g_1 = g_0 = 0\,, \ 
 f_1 = f_2 = 0\,, \ f_3>0$.
In each of the regions $\mathfrak{h}>0$ and $\mathfrak{h}<0$, we can find the following exact solution of (\ref{eqn:WDWeq}) satisfying (\ref{eqn:bc-phi}), 
\begin{equation}
 \Psi(a,\mathfrak{h}) = \left\{ \begin{array}{ll}
		 A \frac{a^c}{\sqrt{\mathfrak{h}}} W_{\kappa,1/4}(w)\,, & \ (\mathfrak{h}>0)\\
		 B \frac{a^c}{\sqrt{-\mathfrak{h}}} W_{\kappa,1/4}(w)\,, & \ (\mathfrak{h}<0)
		  \end{array}\right.\,,
\end{equation}
where $c$ ($>0$), $A$ and $B$ are constants, $\kappa$ and $w$ are defined by
\begin{equation}
 \kappa = -\frac{V}{4\sqrt{f_3}}\left[ c^2 +(p-1)c-2 g_3 \right]\,, \quad
  w = V\sqrt{f_3}\mathfrak{h}^2\,. \label{eqn:defs-kappa-w}
\end{equation}
 and $W_{\mu, \nu}(w)$ is the Whittaker function.

We now require the continuity of $\Psi(a,\mathfrak{h})$ and $\partial_{\mathfrak{h}}\Psi(a,\mathfrak{h})$ at $\mathfrak{h}=0$, which is necessary to ensure the smoothness of the solution. Since 
\begin{align}
\begin{split}
&\lim_{\mathfrak{h}\to +0}\Psi(a,\mathfrak{h}) = A\frac{\pi^{1/2}V^{1//4}f_3^{1/8}}{\Gamma(3/4-\kappa)}\,, \\
&\lim_{\mathfrak{h}\to -0}\Psi(a,\mathfrak{h}) = -B\frac{\pi^{1/2}V^{1//4}f_3^{1/8}}{\Gamma(3/4-\kappa)}\,, 
\end{split}
\end{align}
we have $B=-A$ from the continuity of $\Psi(a,\mathfrak{h})$ at $\mathfrak{h}=0$. 
We also find  
\begin{equation}
 \lim_{\mathfrak{h}\to \pm 0}\partial_{\mathfrak{h}}\Psi(a,\mathfrak{h}) = \mp 2A\frac{\pi^{1/2}V^{3//4}f_3^{3/8}}{\Gamma(1/4-\kappa)}\,. 
\end{equation}
The continuity of $\partial_{\mathfrak{h}}\Psi(a,\mathfrak{h})$ at $\mathfrak{h}=0$ then requires the argument of the Gamma function in the denominator to be $1/4-\kappa = -N$, ($N=0,1,\cdots$). This can be rewritten as 
\begin{equation}
 c^2 + (p-1)c + \left[\frac{\sqrt{f_3}}{V}(4N+1) - 2g_3\right] = 0\,, \
  (N=0,1,\cdots)\,, \label{eqn:eq-c}
\end{equation}
which determines $c$. Therefore we have 
\begin{equation}
 \Psi(a,\mathfrak{h}) = \left\{ \begin{array}{ll}
		 A \frac{a^c}{\sqrt{\mathfrak{h}}} W_{N+1/4,1/4}(w)\,, & \ (\mathfrak{h}>0)\\
		 -A \frac{a^c}{\sqrt{-\mathfrak{h}}} W_{N+1/4,1/4}(w)\,, & \ (\mathfrak{h}<0)
		  \end{array}\right.\,. \label{eqn:z=3sol}
\end{equation}
One can consider a linear combination of solutions with different values of $N$ as far as the corresponding values of $c$ are positive.

For instance, the solution (\ref{eqn:z=3sol}) for $N=0$ corresponds to the ground state and takes the Gaussian form for $\mathfrak{h}$,
\begin{equation}
 \Psi(a,\mathfrak{h}) = A(V\sqrt{f_3})^{1/4} a^c e^{-\frac{V\sqrt{f_3}\mathfrak{h}^2}{2}}\,, 
\end{equation}
which suggests that two-point tensor correlator is completely scale-invariant, $\langle \mathfrak{h}^2\rangle = \mathcal{N}\int d\mathfrak{h}\, \mathfrak{h}^2 \left|\Psi(a,\mathfrak{h})\right|^2 \propto n^{-3}$ since $f_3\propto n^6$, where $\mathcal{N}=(\int d\mathfrak{h}\,  \left|\Psi(a,\mathfrak{h})\right|^2)^{-1}$. This is consistent with the result of \cite{Mukohyama:2009gg}.

In general case of (\ref{eqn:z=3sol}), the expectation value of $w\propto \mathfrak{h}^2$ on $a=const.$ hypersurfaces is independent of $a$ and of order unity, $\langle w \rangle = \mathcal{O}(1)$, for each value of $N$ ($=0,1,\cdots$), provided that $f_3>0$. This again implies that the power spectrum is scale-invariant and independent of $a$ as 
$n^3 \langle \mathfrak{h}^2\rangle = \mathcal{O}(1)\times {\cal M}^2$  
where we have defined the mass scale ${\cal M}$ so that $f_3 \sim n^6/{\cal M}^4$ for large $n$.

\section{General solution near $a=0$}
\label{sec:small-a}

\subsection{Leading order solution  $F_0(\mathfrak{h})$}

Let us now consider the general case for the HL gravity.
By substituting (\ref{eqn:Psi-expanded}) to (\ref{eqn:WDWeq}), at the leading order in $a$, one obtains 
\begin{equation}
 \partial_{\mathfrak{h}}^2F_0 - V^2\left[ f_3\mathfrak{h}^2 + (c+p-1)c - 2 g_3 \right]F_0 = 0\,. 
\end{equation}
By requiring (\ref{eqn:bc-phi}) (with $i=0$) and the continuity of $F_0(\mathfrak{h})$ and $\partial_{\mathfrak{h}}F_0(\mathfrak{h})$, we can easily find (\ref{eqn:eq-c}) determining $c$ and the following solution 
\begin{equation}
 F_0(\mathfrak{h}) = \left\{ \begin{array}{ll}
		 \frac{A}{\sqrt{\mathfrak{h}}} W_{N+1/4,1/4}(w)\,, & \ (\mathfrak{h}>0)\\
		 -\frac{A}{\sqrt{-\mathfrak{h}}} W_{N+1/4,1/4}(w)\,, & \ (\mathfrak{h}<0)
		  \end{array}\right.\,,
\end{equation}
where $N=0,1,\cdots$, provided that $f_3>0$. (One can consider a linear combination of solutions with different values of $N$ as far as the corresponding values of $c$ are positive.) This solution is the same as the previous solution (\ref{eqn:z=3sol}) that was obtained for the strictly $z=3$ case. Hence, for $f_3>0$, in the $a\to +0$ limit we have the scale-invariant and finite power spectrum,
$\lim_{a\to +0}n^3 \langle \mathfrak{h}^2\rangle = 
 \mathcal{O}(1)\times {\cal M}^2$ for large $n$,
which is again consistent with the result of \cite{Mukohyama:2009gg}. 

On the other hand, for $f_3\leq 0$ there is no non-trivial smooth solution satisfying (\ref{eqn:bc-phi}) (with $i=0$). In particular, this is the case in the absence of $z=3$ terms (for which $f_3=0$). This no-go result applies to many gravitational theories (including GR) in which the action does not contain terms with six spatial derivatives.

\subsection{First order correction $F_1(\mathfrak{h})$}

From here, let us consider the higher-order corrections. At the next-to-leading order in $a$, one obtains,
\begin{equation}
 \partial_{\mathfrak{h}}^2F_1 - V^2\left[ f_3\mathfrak{h}^2 + (c+p)(c+1) - 2 g_3 \right]F_1 = 0\,.  \label{eqn:eq-for-F1}
\end{equation}
In each of the regions $\mathfrak{h}>0$ and $\mathfrak{h}<0$, we can easily find the following solution satisfying the boundary condition (\ref{eqn:bc-phi}) (with $i=1$), 
\begin{equation}
 F_1(\mathfrak{h}) = \left\{ \begin{array}{ll}
		 \frac{\tilde{A}}{\sqrt{\mathfrak{h}}} W_{\kappa+\kappa_1,1/4}(w)\,, & \ (\mathfrak{h}>0)\\
		 \frac{\tilde{B}}{\sqrt{-\mathfrak{h}}} W_{\kappa+\kappa_1,1/4}(w)\,, & \ (\mathfrak{h}<0)
		  \end{array}\right.\,,
\end{equation}
where $\tilde{A}$ and $\tilde{B}$ are constants, $\kappa$ and $w$ are defined in (\ref{eqn:defs-kappa-w}), 
$\kappa_1 = - V (2c + p)/(4\sqrt{f_3})$
and $c$ has already been determined by (\ref{eqn:eq-c}).

By requiring the continuity of $F_1(\mathfrak{h})$ at $\mathfrak{h}=0$, one obtains $\tilde{B}=-\tilde{A}$. The continuity of $\partial_{\mathfrak{h}}F_0(\mathfrak{h})$ at $\mathfrak{h}=0$ then requires that $1/4-\kappa-\kappa_1$ be a non-negative integer or that $\tilde{A}=0$. Since $1/4-\kappa$ is already set to be a non-negative integer, the former condition can be satisfied only if $\kappa_1$ is an integer, which requires a fine-tuning. Avoiding the fine-tuning, we conclude that $\tilde{A}=0$, i.e. $F_1(\mathfrak{h}) = 0$.

\subsection{Second order correction $F_2(\mathfrak{h})$}

At the next-to-next-to-leading order in $a$, one obtains 
\begin{align}
&\partial_{\mathfrak{h}}^2F_2 - V^2\left[ f_3\mathfrak{h}^2 + (c+p+1)(c+2) - 2 g_3 \right]F_2 \notag \\
&= V^2(f_2\mathfrak{h}^2-6g_2)F_0\,. 
\end{align}
Unlike its counter part (\ref{eqn:eq-for-F1}) for $F_1$, this equation for $F_2$ is sourced by $F_0$. Therefore, $F_2$ cannot be identically zero. Once $F_0(\mathfrak{h})$ is given, $F_2$ is determined by this equation and (\ref{eqn:bc-phi}) (with $i=2$). While the computation is straightforward, the result for $F_2$ is complicated. Hence we show the structure of the solution for $N=0,1,2$ without explicit expressions. Using the leading-order solution,
\begin{equation}
 F_0(\mathfrak{h}) = \left\{
		      \begin{array}{ll}
		       A_0 \exp\left(-\frac{1}{2}w\right)\,, & (N=0)\\
		       (1-2w) A_1 \exp\left(-\frac{1}{2}w\right)\,, & (N=1)\\
		       \left(1-4w + \frac{4}{3}w^2\right) A_2 \exp\left(-\frac{1}{2}w\right)\,, & (N=2)
		      \end{array}
\right.\,,
\end{equation}
$F_2$ is shown to have the form
\begin{equation}
 F_2(\mathfrak{h}) = \sum_{\tilde{N}=0}^{N+1}a_{N,\tilde{N}}w^{\tilde{N}}A_N\exp\left(-\frac{1}{2}w\right)\,,
\end{equation}
where $w=V\sqrt{f_3}\mathfrak{h}^2$. For each $N$, $c$ is determined by (\ref{eqn:eq-c}), $A_N$ is an integration constant, $a_{N,\tilde{N}}$ ($\tilde{N}=0,\cdots,N+1$) are constants determined by the parameters in (\ref{eqn:WDWeq}).

\section{Discussions}

We have shown that, in many theories of gravity including general relativity, the Wheeler-DeWitt equation with the DeWitt boundary condition does not admit a wave function of the universe that gives non-divergent correlation functions of $h$ on $a=const.$ hypersurfaces near the classical big-bang singularity once tensor perturbations around a homogeneous and isotropic closed universe are taken into account, where $a$ is the scale factor of the universe and $h$ is the amplitude of tensor perturbation. The correlation functions of perturbations such as the power spectrum diverge near the classical big-bang singularity.

On the contrary, we have shown that the Ho\v{r}ava-Lifshitz (HL) gravity provides a satisfactory DeWitt wave function when tensor perturbations are included. In the case of the strict $z=3$ anisotropic scaling, we have analytically given the exact DeWitt wave function. Furthermore, in more general cases with relevant deformations, we have analytically obtained the DeWitt wave function near the classical big-bang singularity up to the second-order in the scale factor. These DeWitt wave functions in the HL gravity are uniquely determined by the parameters in the action, the operator ordering and the quantum number $N$ parameterizing the ground ($N=0$) and excited ($N=1,\cdots$) states of the perturbations. As a consistency check, we have shown that the DeWitt wave function in the HL gravity correctly reproduces the scale-invariant power spectrum of perturbations that was found previously in \cite{Mukohyama:2009gg}.

We have restricted our consideration to only one mode of the tensor perturbations after the harmonic expansion (\ref{eq:perturbation}). If we take into account other modes and if impose the boundary condition (\ref{eqn:bc-phi}) for all of them then the algebraic equation (\ref{eqn:eq-c}) determining $c$ will be modified. By demanding the positivity of $c$, we then obtain a theoretical constraint on the operator ordering parameter $p$. It is certainly worthwhile studying this issue in more detail in future work.

It is known that the wave function of the universe in homogeneous and anisotropic models tends to vanish towards the classical big-bang singularity (see e.g.~\cite{Kleinschmidt:2009cv}). It is also known that a Bianchi IX spacetime in the small anisotropy limit corresponds to a closed FLRW spacetime with a particular mode of tensor perturbation. Therefore our result in GR suggests that in the Bianchi IX minisuperspace the wave function should spread over the space of anisotropies, i.e. the anisotropies are not suppressed~\footnote{We thank an anonymous referee for pointing this out.}. Indeed, eq.~(8) of \cite{Kleinschmidt:2009cv} would correspond to a complex value of $c$ and (\ref{eqn:GR-leading-a}) still suggests that the tensor mode corresponding to homogeneous anisotropy is not suppressed. Furthermore, nonlinear completions of general tensor modes, forming the full set of gravitational degrees of freedom without any symmetries, do not fit into the Bianchi IX minisuperspace and are also unsuppressed, rendering the description based on GR broken at the classical big-bang singularity.

In this Letter we have focused on the analytical investigation of the DeWitt wave function in vacuum. We leave the numerical estimation, the effect of the inclusion of matter fields and the detailed discussion of the interpretation of the cosmic wave function for future work.

\medskip
{\it Acknowledgments.} 
The work of H.M. was supported by JSPS KAKENHI Grant No. JP22J01284.
S.M.'s work was supported in part by Japan Society for the Promotion of Science Grants-in-Aid for Scientific Research No. 17H02890, No. 17H06359, and by World Premier International Research Center Initiative, MEXT, Japan. 
The work of A.N. was supported in part by JSPS KAKENHI Grant Numbers 19H01891 and 20H05852.

\end{document}